# Direct and real-time analysis in a plasma reactor using a compact FTICR/MS: degradation of acetone in nitrogen and by-products formation

Sébastien Thomas[a], Nicole Blin-Simiand[c], Michel Héninger[b,a], Pascal Jeanney[c], Joel Lemaire[a], Lionel Magne[c], Hélène Mestdagh[a], Stéphane Pasquiers[c], Essyllt Louarn[a,*,†]

a- Université Paris-Saclay, CNRS, Institut de Chimie Physique, 91405, Orsay, France.

b- AlyXan, 91260 Juvisy-sur-Orge, France

c- Université Paris-Saclay, CNRS, Laboratoire de Physique des Gaz et des Plasmas, 91405, Orsay, France.

**ABSTRACT:** Methods for reduction of Volatile Organic Compounds (VOCs) content in air depend on the application considered. For low concentration and low flux, non-thermal plasma methods are often considered as efficient. However, the complex chemistry involved is still not well understood as there is a lack of datasets of byproducts formation. So as to overcome this issue, rapid analytical methods are needed. We present the coupling of a rapid chemical ionization mass spectrometer (CIMS) for the real-time analysis of the VOCs formed during a degradation experiment. The high resolution instrument used allows for chemical ionization and direct quantification of non-targeted compounds. We present degradation experiments of acetone in a photo-triggered nitrogen plasma discharge. Two regimes were highlighted: efficient conversion at low concentrations (<100ppm) and moderate efficiency conversion at higher concentrations (>100ppm). Those two regimes were clearly delimited as the sum of two exponential curves occuring at respectively low and high concentrations. Many by-products were detected, in particular HCN presented a significantly high yield. Nitrile compounds (acetonitrile, propionitrile,...) are formed as well. To a lower extent, ketene, acetaldehyde and formaldehyde are observed. The association of the high resolution mass spectrometer to the plasma reactor will allow further insights into the plasma chemistry and comparison to modelisation.

[**]corresponding author: essyllt.louarn@u-psud.fr

† Present address: Université de Lyon, Université Claude Bernard Lyon 1, CNRS, IRCELYON, 69269 Villeurbanne, France

Air pollution is a major health issue worldwide. World Health Organization estimated that, in 2016, 4.6 million people are subjected to premature death due to bad air quality from cardiovascular and pulmonary diseases. Acute pollution such as haze episodes in the cities tends to present pollution as occasional. However, high level of pollutants is also chronic and quite common in urban environments, outdoor as well as indoor. Moreover, pollutants can be transformed in the environment and their combinations are responsible for the formation of other noxious species, such as ozone or ultra-fine micro-particles. Volatile organic compounds (VOCs) are among those pollutants. First, part of the VOCs is toxic by themselves, for instance benzene and formaldehyde are regulated in the different environments. Second, in highly polluted cities, the combination of VOCs and NOx produce a reaction triggered by light that creates ozone, a highly toxic pollutant often observed during summer and long sunny periods in many cities around the world[1–3]. Moreover, VOCs are known to promote the formation of secondary organic aerosols, leading to small particles of a few micrometer size or smaller[4–6]. Those microparticles, after inhalation by a human being, may enter the lungs and go deep in the alveoli increasing lung cancer probability[7].

In the EU and the US, recent regulations limited VOCs emission for vehicles, and succeeded in decreasing VOCs concentration due to transport[8]. Hence, McDonald et al[9] identified that nowadays volatile organic compounds are mainly emitted from households themselves, for instance from coatings, paintings and diverse care-products. As people in urban areas spend most of their time indoor, there is a growing interest in the indoor air quality.

Many solutions have been proposed to decrease VOCs in indoor air[10] as well as emissions by industries and transports. Many of them are available for the public. Those solutions, alone or associated to each other, are based on very different techniques, such as plasma degradation[11,12], photocatalysis[13], biofilters[14], High Efficiency Particulate Air (HEPA) filters, thermal degradation,.... Contrary to the other techniques, non-thermal plasmas are efficient in a wide range of concentration from below 100 ppb to more than 1000 ppm[11]. It makes it an interesting technique for air depollution at low emission levels such as in indoor air depollution[15] or odor removal[16].

Electrical discharges, such as corona discharges, dielectric barrier discharges (DBD), or the photo-triggered discharge (PTD) used in this study, work at room temperature and allow rapid on-off switching at moderate energy. Atomic and/or excited nitrogen and oxygen, produced in the air plasma, are used as precursors for the initial cleavage reactions of the VOC molecule. It gives rise to a cascade of chemical reactions involving various radical species.

In the way down to total degradation, a wide range of chemical byproducts are formed, some of them being toxic, and have to be analyzed with the best possible accuracy. When monitored, they are usually evaluated by GC-FID or GC-MS, or FTIR[17–19]. For example, the degradation of acetaldehyde[17,20], acetone[18,21] or toluene[22] in nitrogen or in nitrogen/oxygen mixture plasmas produced in DBD or in PTD have been studied. Many organic byproducts have been identified among which hydrogen cyanide, hydrocarbons $CH_4$, $C_2H_{y=2,4,6}$ in nitrogen plasma, and oxygenated compounds such as formaldehyde or acetaldehyde in air. GC measurement delay is typically of a few minutes at best. However, a better time resolution would be very useful to draw a more complete picture of the processes going on in the gas mixture and to have a better understanding of the complex plasma chemistry. In order to measure the concentrations of the large number of compounds present in a plasma discharge reactor used for the degradation of a VOC a complementary method is necessary. Then to obtain a direct glimpse of the kinetic processes occurring in the plasma, it should analyze the mixture every few seconds, simultaneously for all compounds.

Chemical Ionization (CI) is a mass spectrometry ionization technique that enables direct and real-time analysis of gas mixtures. It is based on the reaction of a known ion, called precursor, with an analyte through an ion-molecule reaction. When chosen carefully, the precursor leads to a unique pseudo-molecular ion for each molecule of the mixture. Different families of compounds may be targeted according to the precursor used. One of the earliest CI methods uses precursor ions formed from $CH_4$[23]. The $CH_5^+$ and $C_2H_5^+$ resulting ions react with any kind of organic compounds, at the cost of extensive fragmentation. In the environmental chemistry field, the most common precursor is $H_3O^+$, formed from $H_2O$. In that case, precursor formation and ion-molecule reaction with the analytes may be managed in an ion source where temperature, pressure and reaction time are controlled. The resulting technique, often called "Proton Transfer Reaction Mass Spectrometry" (PTR-MS)[24], is therefore quantitative[24,25]. The $H_3O^+$ precursor has many advantages: it doesn't react with the major constituents of air ($N_2$, $O_2$, Ar, $CO_2$, etc...); organic species such as oxygenated VOCs form a convenient $MH^+$ ion and fragmentation is limited[25]. Though, some molecules do not react with $H_3O^+$ as they have lower proton affinities, in particular small alkanes and ethylene. As alkanes and alkenes are supposed to be byproducts of acetone degradation, use of other precursors is needed. Due to humidity interactions, many precursors are forbidden, such as those formed from i-$C_3H_8$, $CH_4$ or $CF_4$[26]. On the contrary, $O_2^+$ reacts by charge transfer with a large number of VOCs[27–30] including ethane and $NO_X$. The drawback is partial fragmentation in many cases. Finally, high-resolution mass spectrometry meets the need of the challenging direct analysis of a complex mixture without separation, as identification through molecular formula is made possible. Fourier Transform Ion Cyclotron Resonance Mass Spectrometers (FTICR/MS) are known for their very high resolution. However, use of a superconductor magnet makes these instruments non-transportable and expensive. Recently, Heninger *et al*[31] presented a compact FTICR/MS using a permanent magnet whose results allowed for rapid detection (a few seconds) and identification of targeted and untargeted VOCs. Internal design and configuration of the ICR cell[35] results in easy implementation of quantitative chemical ionization with numerous precursors.

In this work, we present an innovative coupling of the CI-FTICR/MS using multiple precursors such as $H_3O^+$ and $O_2^+$ with a photo-triggered discharge reactor designed for kinetic studies on removal of VOCs in plasmas of atmospheric gases. This system is used for analysis in real-time of the gas mixture formed from acetone degradation reactivity. To demonstrate the coupling efficiency, we present first the comparison with GC measurements, then real-time results of acetone degradation and by-products formation, and, finally, its relevance for the understanding of acetone kinetics in the homogeneous plasma produced in $N_2/CH_3COCH_3$ mixtures. The experimental data obtained will be of importance for further validating a detailed kinetic scheme developed for plasma chemistry models.

## MATERIALS AND METHODS

### Gas mix production

Acetone is available as a gas diluted in nitrogen at 1000 ppm (Crystal mix, B50 Alphagaz, AirLiquide). Pure nitrogen bottle (B50, Alphagaz 1) is purchased from Air Liquide. The pressure in the total volume of the photo-triggered discharge reactor, $V_T$ = 8.85 L, is measured by an MKS baratron® gauge (MKS Instrument France, Le Bourget, France). Different concentrations of acetone are operated in the reactor by pressure adjustment for each gas addition until a total pressure of 0.46 bar, value chosen in order to ensure the homogeneity of the plasma from one current pulse to the next[17,32] (see also below). Before each mix formation, the volume $V_T$ is emptied using a turbo-molecular pump at a residual pressure well below 0.1 mbar. First, gas is introduced from the 1000 ppm acetone bottle at a specific pressure depending on the final wanted acetone partial pressure. Afterwards, pure nitrogen gas is added up to 0.46 bar.

### Plasma reactor

The operating principle of the photo-triggered discharge 'UV510' has been extensively described elsewhere[17,32]. It is a kind of pre-ionized discharge allowing the production of a perfectly homogeneous non-thermal plasma between two metallic electrodes. The plasma reactivity proceeds during a short current pulse of 60 ns (electron collisions on molecules to produce atoms, radicals, excited states, ions and various reactions between these species). The discharge volume is $V_D$=50 cm$^3$, enclosed in a stainless-steel cell with a volume $V_C$=500 cm$^3$ (Figure 1). A compressor is used to produce a gas flow (7 L/s at 0.46 bar) through the inter-electrode space, in a closed loop with total volume $V_T$ much higher than $V_D$, and the pulse repetition frequency (1 Hz) is chosen such that $V_C$ is completely renewed between two consecutive pulses. After each current pulse follows a long post-discharge period, during which the chemical reactivity between neutral species develops. The primary organic molecules progressively disappear in the gas mixture whereas stable by-products accumulate in $V_T$ as the mixture undergoes an increasing number of current pulses, each for equal deposited electrical energy in $V_D$. Concentrations of molecules are measured in $V_T$ as a function of the number of pulses, noted $N_D$ (discharge number) in the text and figures. For the present experiment the deposited energy in the plasma volume, per current pulse, is 4.6 J/pulse (92 J.L$^{-1}$). Moreover, the number $N_D$ is equivalent to 1 s.

### FTICR/MS instrument

BT4 is a compact FTICR/MS instrument (AlyXan, Juvisy-sur-Orge, France)[31]. The instrument is based on a Halbach permanent magnet[33,34], and allows for chemical ionization (CI) in the ICR cell as it makes use of a sequential introduction system. Typical analysis gives a complete mass spectrum (0-300 u) in 1 to 4 s[31]. Different precursor ions may be used in positive mode ($H_3O^+$[25,35,36], $O_2^+$, $CF_3^+$[26], $C_6H_4F_2.H^+$[37], etc...) and negative mode[38] ($HO^-$, $O^-$, etc...). As the ICR trap works sequentially, it is possible to switch during the same experiment between different chemical ionization precursors, and even between CI and electronic ionization (EI)[31]. However, this switch comes at the cost of time response.

For this study, it was decided to maintain only one CI precursor per experiment to obtain the best time resolution possible, as the discharge has a frequency of 1 Hz. Two different CI precursors were used: $H_3O^+$ and $O_2^+$. $H_3O^+$ is a generalist precursor whose reactivity allows for identification of a large number of oxygenated or nitrogenated VOCs. It reacts by proton transfer and forms mostly one ion product ($MH^+$) whose mass is 1.008 amu above the mass of the neutral molecule.

$H_3O^+$ is formed from the interaction of 40 eV accelerated electrons with a pulse of water vapor (P=10$^{-6}$ mbar). After circa 300 ms, no $H_2O^+$ is observed and only $H_3O^+$ is detected. $O_2^+$ is formed similarly using pure $O_2$.

As chemical ionization is performed in controlled pressure and temperature conditions with a user-set reaction time, direct quantification is possible when knowing the reaction rate coefficient and the signal height of the precursor ion and daughter ions[25]. However, due to collision effects on precursor ions with the molecule of the matrix gas, the concentrations obtained have to be corrected. This factor is the same for all compounds and has been determined to be 1.50 using calibrated gas. Introducing known concentrations of acetone in the reactor, we found a factor of 1.47 in good agreement (Supplementary S1). All acetone concentrations are calculated taking that value into account.

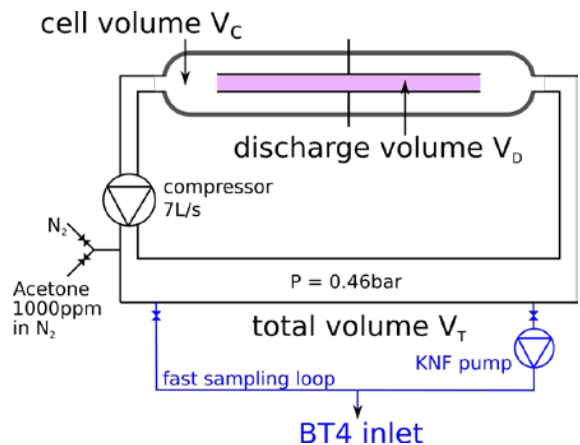

**Figure 1.** Scheme of the experiment. Discharge volume (purple) is $V_D$=50 cm³; $V_C$=500 cm³; $V_T$=8.85 L. The additional sampling loop for BT4 apparatus is in blue. For complete description see text.

Association of the reactor and BT4

Figure 1 presents the association of the BT4 instrument and the plasma reactor. BT4 was adapted to the reactor through its "sniffer" inlet[34]. This inlet enables direct introduction of a gas from atmospheric pressure into the apparatus. The pressure drop is produced in two steps. The first drop is from atmospheric pressure down to a pressure of a few millibars using a needle valve and a membrane primary pump with a flow of 10 to 50 mL/min connected to a waste. The second drop uses a stainless-steel capillary (internal diameter, 130 µm; length 50 mm) placed before the three-way valve. To connect the inlet to the plasma reactor, a fast sampling loop was added to the main gas loop as an intermediate between the reactor and the instrument. A PTFE-coated mini-diaphragm pump (laboport N86KT, KNF Neuberger SAS, Village-Neuf, France) provided a *circa* 6 L/min flow in the 1/8 inch PFA tubing. A 1/8 inch tee connected this loop to the entrance of the sniffer on the instrument. The response time was a few seconds.

BT4 sampling through the sniffer had a small, yet observed, influence on the tota pressure of the reactor. A steady decrease of the pressure was observed. After 20min, the pressure in the main gas loop was weakened by a maximum of 10%. It was then chosen to limit all experiments to 20min, *i.e.* 1200 discharges.

### Gas chromatography analysis

During this study, use of gas chromatography widened the range of species observed and allowed for comparison to the FTICR measurements. The Gas Chromatograph was equipped with a Flame Ionization Detector (GC-FID; 7890A Agilent Technologies) and a split/splitless injector. Two separated columns were available in the GC: a Varian CP-PoraBOND-Q (25 m, 0.25 mm, 3 µm), for alkane analysis and a Varian Al2O3 (50 m, 0.32 mm, 5 µm) for oxygenated VOCs analysis. Each column had its separated injector, they were set to T=200°C for both columns. Oven initial temperature was set to 60°C during 5 minutes for both columns, then temperature ramps of 25°C.min$^{-1}$ up to 130°C and 30°C.min$^{-1}$ up to 120°C were applied respectively to the Porabond Q and the Al2O3 columns. The vector gas is helium at 0.5mL/min. The detector temperature was set at 300°C.

For a given acetone concentration, after a chosen number of discharges $N_D$, samples of 1 cm$^3$ were pumped from the gas mixture by a Hamilton gastight syringe. Each sampling is done twice to analyze on both columns. This allows to investigate the global composition of the mixture. For each analysis, the experiment is renewed (filling, discharge, sampling). This technique gave good results for previous experiments on propane[39], acetaldehyde[17], and acetone[40].

To be able to compare to BT4 (degradation speed and byproducts), GC measurements were conducted on a 500 ppm acetone initial concentration for 0, 125, 375 and 500 discharges.

## COMPARISON OF GC-FID AND BT4 RESULTS

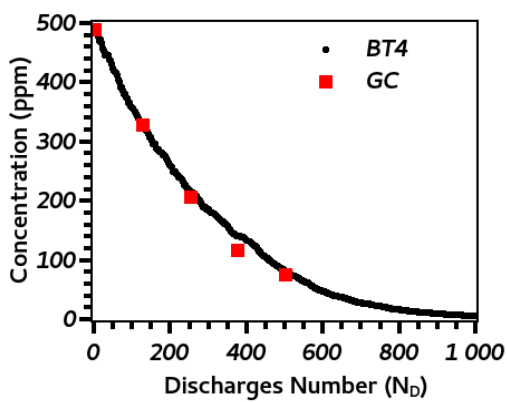

**Figure 2.** Acetone degradation monitored in real-time by BT4 (black dots) and GC (red squares). Acetone concentrations are presented in ppm and the X axis is $N_D$, the number of discharges, which is equivalent to a time (in seconds).

Figure 2 presents concentration values obtained by BT4 and GC-FID during acetone degradation (see also Supplementary S2). The discharge frequency is 1 Hz, therefore the number of discharges is equivalent to a degradation time in seconds. An exponential-like decrease of acetone concentration is observed. Acetone real-time concentrations as obtained by BT4 are comparable to the values obtained by GC-FID. It has to be noted that the BT4 values are obtained during one experiment, when GC-FID results are obtained from 5 different experiments. BT4 is then efficient to provide real-time analytical results comparable to GC data.

## RESULTS OF THE DEGRADATION EXPERIMENT

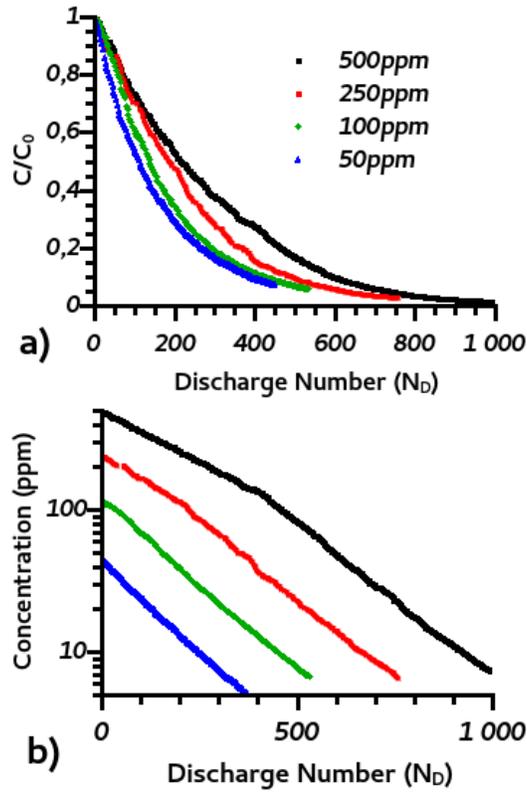

**Figure 3.** Evolution of acetone concentration during degradation in the reactor in pure nitrogen for different initial concentrations between 500 ppm and 50 ppm (a) in relative concentration ($C/C_0$), and (b) in log scale. Discharges have a frequency of 1 Hz.

**Acetone degradation**

Acetone degradation was monitored at different initial concentrations: 500 ppm, 250 ppm, 100 ppm and 50 ppm. Concentrations of acetone at the beginning of each experiment were measured using BT4 calibration and were respectively: 488, 239, 114 and 45 ppm. The maximum deviation of the initial acetone concentration from GC values is 10%. Acetone concentration was monitored until it is well below 10 ppm.

Figure 3 presents the real-time evolution of the acetone concentration during the degradation process. For the four studied initial concentrations, acetone concentration shows an exponential-like decrease (Figure 3-a). The characteristic number of discharges $N_C$ is respectively equal to 306±1, 265±2, 184±1 and 173±1. $N_C$ represents the inverse of the slope at initial time:

$$\frac{dC}{dt}(t=0) = C_0\left(1 - \frac{N_D}{N_C}\right) \qquad (1)$$

where $C_0$ is the initial concentration. Furthermore, the initial quantity of acetone converted in the active discharge volume $V_D$ for one pulse, noted $C_D$, is calculated according to:

$$C_D = C_0\left(1 - exp\left(\frac{-1}{N_C}\right)\right)\frac{V_T}{V_D} \qquad (2)$$

where $V_T$ and $V_D$ are respectively the total volume of the plasma reactor (8.85 L) and the volume of the discharge (50 mL) and $V_T/V_D$ represents a diluting factor. The values of $C_D$ are 305±3, 173±18, 118±12 and 50±5 ppm/pulse for the initial concentrations respectively equal to 500, 250, 100 and 50 ppm. For initial concentrations of 100 ppm and lower, $C_D$ is equal to $C_0$; the acetone molecules present in the plasma volume $V_D$ are then entirely converted with one current pulse. On the contrary, for higher initial concentrations, $C_D$ is inferior to $C_0$; the conversion with one current pulse is therefore not total; it drops from 66% of the molecules converted at 250 ppm to 57% at 500 ppm. The efficiency of acetone removal (defined by percentage of concentration removed) by the photo-triggered discharge is then greater for low concentrations than for high concentrations. This feature observed with BT4 corroborates previous results obtained using GC-FID[21].

In Figure 3-b, for the two lowest initial concentrations (50 and 100 ppm), the curves display one slope, as expected from a reactor where conversion of the acetone in the plasma volume is total for each pulse. On the contrary, for higher concentrations, we observe a curvature separating two different lines with two different slopes. This is quite unexpected since we would have presumed a continuous deviation from exponential decrease of acetone concentration, represented as a curved line in the log scale.

The high concentration acetone degradation presents then a two steps process, each one defining a straight line on the curves in Figure 3-b. The first step, for concentrations superior to circa 100 ppm, presents a constant slope, which indicates a different behavior from low concentrations, due to partial depletion of the acetone and accumulation of degradation products. In the second step, the slope appears close to the one observed for the lowest concentrations (*i.e.* $C_0$ = 50 and 100 ppm).

In that second part of the degradation curves, $N_C$ is equal to 198±1 and 202±1 at, respectively, 250 ppm and 500 ppm initial concentrations. Those values are slightly higher than the $N_C$ obtained for the lowest initial concentrations (for instance at $C_0$ = 50 ppm, $N_C$ is 174) suggesting a slightly lower efficiency of the degradation process during this second step.

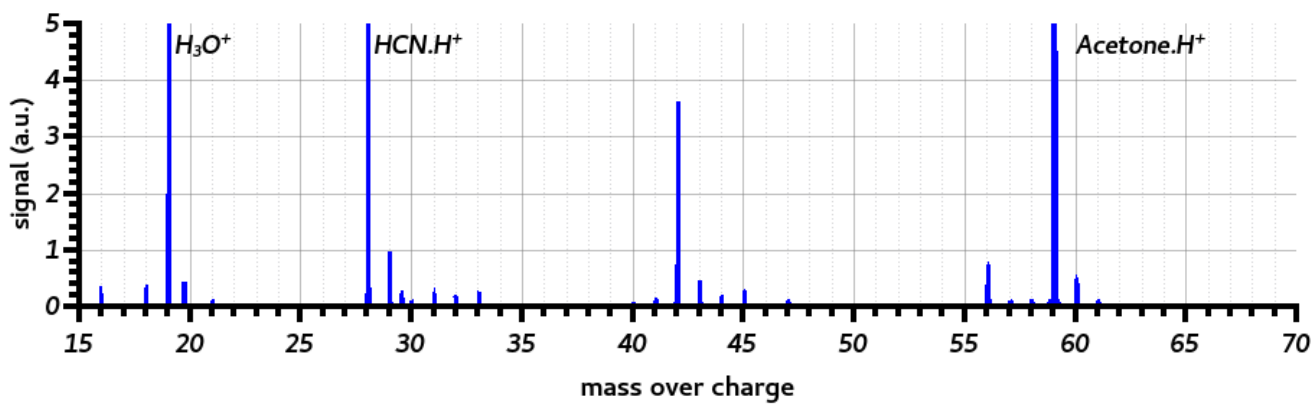

**Figure 4.** Mass spectrum (CI/$H_2O$) of the degradation of 500 ppm acetone in nitrogen after 375 current pulses. The spectrum is zoomed to lower signals below 5; $H_3O^+$, $HCN.H^+$ and $Acetone.H^+$ signals are truncated.

**Degradation products and their evolution**

Many by-products are observed during the degradation of acetone in $N_2$ gas (Figure 4). Main by-products are nitrile compounds: HCN ($MH^+$ 28.0187u), $CH_3CN$ ($MH^+$ 42.0344u) and $C_2H_5CN$ ($MH^+$ 56.0500u). Lower concentrations of other nitrile compounds are detected: $HC_3N$ ($MH^+$ 52.0187u) and $C_2H_3CN$ ($MH^+$ 54.0344u). Some oxygenated species are also observed. Two aldehydes are detected: formaldehyde ($CH_2O$: $MH^+$ 31.0184u) and acetaldehyde ($CH_3CHO$: $MH^+$ 45.0340u); methanol was also identified ($CH_3OH$: $MH^+$ 33.0340u) and ketene $H_2C=C=O$ ($MH^+$ 43.0184u).

Use of $O_2^+$ precursor allowed to identify other byproducts such as alkanes. $O_2^+$ reactivity involves ionization of a large family of compounds and often many fragments are observed. Acetone is identified as two main ions: m/z 58.0419 ($M^{+\bullet}$) and a fragment $[M-CH_3]^+$ at m/z 43.0184. Methane, HCN and $CH_3CN$ cannot be detected with $O_2^+$ precursor as their IE is higher than the one of $O_2$ (respectively 12.6eV, 13.6 eV and 12.2 eV). However, $O_2^+$ brings information on the presence of hydrocarbons (for instance ethene, propane and higher). Propane is observed at m/z 44.0626. High resolution is necessary to be able to take apart the signal of the $^{13}C$ isotope of the m/z 43 fragment of acetone (m/z 44.0212) and propane signal. At m/z 44, three reacting species are identified (Figure 5): propane ($M^{+\bullet}$ m/z 44.0626), $^{13}C^{12}CH_3O^+$ first isotope of acetone fragment (m/z 44.0212) and acetaldehyde ($M^{+\bullet}$ m/z 44.0262). Measured masses are respectively 44.0644 u, 44.0226 u and 44.0287 u. The precision is below 0.5 $10^{-3}$ u and athe ccuracy is 0.9 $10^{-3}$ u for the highest peak (isotope) and 2.5 $10^{-3}$ u for acetaldehyde (Supplementary S3).

Real-time evolution of the concentrations of those species is presented in Figure 6. All nitrile compounds and methanol present similar evolution patterns: a maximum is observed for a discharge number of 450. All those byproducts present a steady decrease afterwards. HCN achieves a maximum concentration of 280 ppm for an initial concentration of 500 ppm of acetone, which represents 18.7% of the total possible carbonated species from the initial acetone concentration. All the byproducts are eventually destroyed by the plasma reactor given sufficient time. Carbonyl compounds such as formaldehyde, acetaldehyde and ketene follow a slightly different regime as the maximum is obtained much sooner, after 200 discharges. Formation of those oxygenated species may be linked to different processes.

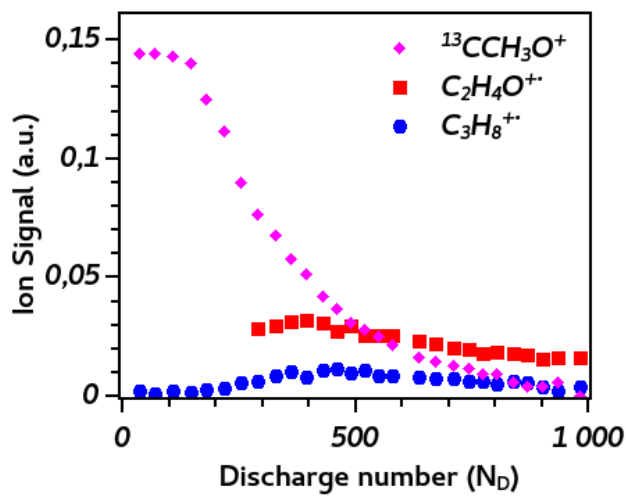

**Figure 5.** Evolution of three isobaric compounds signals as observed in real-time at m/z 44 ($O_2^+$ precursor, 500 ppm): First $^{13}C$ isotope of acetone fragment $C_2H_3O^+$ (magenta diamond), acetaldehyde $C_2H_4O^{+\bullet}$ (red square) and propane $C_3H_8^{+\bullet}$ (blue dot).

### INSIGHTS INTO PLASMA REACTIVITY

On the real-time data of acetone degradation, we observed a rapid slope change in concentration evolution, delimiting two different regimes: a high conversion regime in the plasma volume (~100% conversion) for low concentrations (typically less than 100ppm), and a low conversion regime for high concentrations (> 100ppm). The latter regime was highly dependent on the initial concentration, whereas the first one was only slightly dependent on it. The high concentration regime can be explained by a stoichiometric effect, presenting an excess of the sum acetone + byproducts compared to the reactive species. After degradation to the ultimate products, following several hundreds of discharges, stoichiometry is inversed and leads to a regime where reactive species are in excess.

All the more, BT4 brought new information on the products formed during the degradation process. Previous GC-FID studies allowed to identify CO, $CH_4$, $C_2H_6$ and $H_2$ as major products[21], MS analysis further added HCN to the list. The two techniques presented compatible and, more importantly, complementary results.

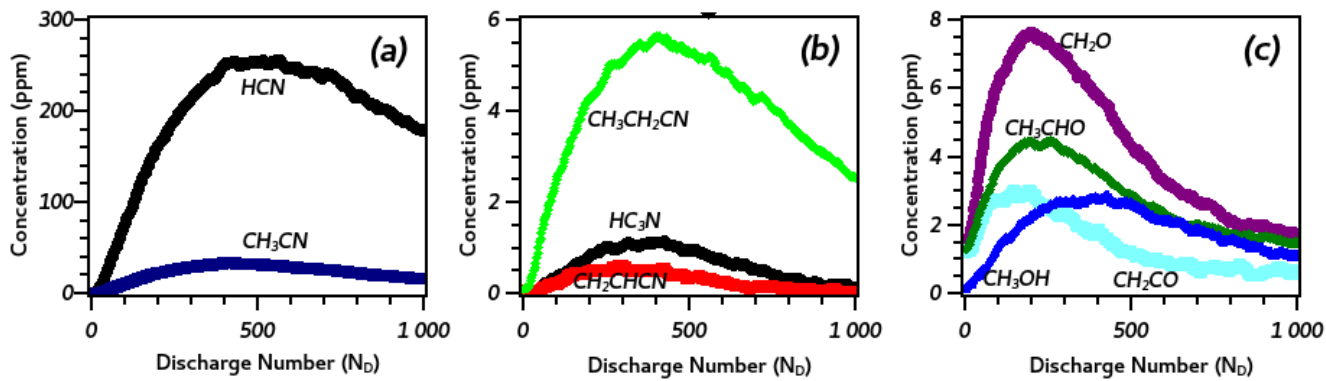

**Figure 6.** Concentration of the byproducts observed during acetone degradation (500 ppm in pure N2); (a) nitric acid and acetonitrile; (b) propionitrile, acrylonitrile and propiolonitrile; (c) methanol, formaldehyde, acetaldehyde and ketene.

The observed degradation products presented a production-degradation profile. At first, they are issued from the degradation of acetone, but afterwards they are also destroyed by the plasma and/or their production is reduced due to the removal of upward species involved in the kinetic chain. Their concentrations decrease steadily at high discharge number, most probably owing to the increase of removal effect.

Acetone fragmentation is mainly governed by its quenching of the nitrogen metastable states. Previous studies on the UV510 reactor have suggested that it leads to the production of H, $CH_3$, and $CH_3CO$[21], the exit route given in (3) being also observed in UV-photodissociation[41,42]:

$N_2^* + CH_3COCH_3 \rightarrow CH_3CO + CH_3 + N_2$ (3)

$N_2^* + CH_3COCH_3 \rightarrow CH_3COCH_2 + H + N_2$ (4)

Afterwards numerous compounds are formed in the mixture owing to radicals' reactivity, but the direct formation of $H_2$, $CH_4$, and CO by the quenching processes seems also plausible if one refers to studies about other VOCs[17,21].

Recombination of the $CH_3$ radical with N radical would lead to HCN and $H_2$ formation by:

$CH_3 + N \rightarrow HCN + H_2$ (5)

Moreover, $CH_3$ recombines into longer aliphatic chains and then with a CN radical or with another CN-containing species, giving rise to the nitrile compounds series ($CH_3CN$, $C_2H_5CN$, $C_2H_3CN$ and $HC_3N$) observed in our study.

Recombination of $CH_3CO$ radical may explain the formation of the ketene:

$H + CH_3CO \rightarrow CH_2CO + H_2$ (6)

Dihydrogen is formed as well. Many other recombination processes are leading to $H_2$ formation.

To a lower extent, some formaldehyde is observed in the mixture. It may be explained by:

$H + CH_3CO \rightarrow CH_3 + HCO$ (7)

$HCO + HCO \rightarrow CH_2O + CO$ (8)

Finally, the coupling of Btrap instrument and the photo-triggered discharge reactor give a new insight of the chemistry occurring in a cold plasma. The real-time results are of great interest for an overall understanding of molecule kinetics in such a reactive medium. New reactivity paths, not specified before, can be highlighted following comparison between predictions of numerical models and measurements. In particular, the measurement of the HCN concentration profiles at different acetone initial concentrations should be of importance for the understanding of the acetone fragmentation process leading to production of the methyl radical and its recombination. Future works will attempt to resolve this issue.

ACKNOWLEDGMENT

The author would like to thank Gérard Mauclaire and AlyXan's team. IDEX Paris Saclay funded the PhD salary of Sébastien Thomas under the interdisciplinary PhD program.

REFERENCES

(1) Haagen-Smit, A. J. *Ind. Eng. Chem.* **1952**, *44* (6), 1342–1346.

(2) Pusede, S. E.; Steiner, A. L.; Cohen, R. C. *Chem. Rev.* **2015**, *115* (10), 3898–3918.

(3) Ding, A. J.; Wang, T.; Thouret, V.; Cammas, J.; Nédélec, P. **2008**.

(4) Turpin, B. J.; Huntzicker, J. J. *Atmospheric Environment* **1995**, *29* (23), 3527–3544.

(5) Carter, W. P. L. *Atmospheric Environment. Part A. General Topics* **1990**, *24* (3), 481–518.

(6) Pandis, S. N.; Harley, R. A.; Cass, G. R.; Seinfeld, J. H. *Atmospheric Environment. Part A. General Topics* **1992**, *26* (13), 2269–2282.

(7) Hamra Ghassan B.; Guha Neela; Cohen Aaron; Laden Francine; Raaschou-Nielsen Ole; Samet Jonathan M.; Vineis Paolo; Forastiere Francesco; Saldiva Paulo; Yorifuji Takashi; et al. *Environmental Health Perspectives* **2014**, *122* (9), 906–911.

(8) May, A. A.; Nguyen, N. T.; Presto, A. A.; Gordon, T. D.; Lipsky, E. M.; Karve, M.; Gutierrez, A.; Robertson, W. H.; Zhang, M.; Brandow, C.; et al. *Atmospheric Environment* **2014**, *88*, 247–260.

(9) McDonald, B. C.; Gouw, J. A. de; Gilman, J. B.; Jathar, S. H.; Akherati, A.; Cappa, C. D.; Jimenez, J. L.; Lee-Taylor, J.; Hayes, P. L.; McKeen, S. A.; et al. *Science* **2018**, *359* (6377), 760–764.

(10) Luengas, A.; Barona, A.; Hort, C.; Gallastegui, G.; Platel, V.; Elias, A. *Rev Environ Sci Biotechnol* **2015**, *14* (3), 499–522. https://doi.org/10.1007/s11157-015-9363-9.

(11) Schiavon, M.; Torretta, V.; Casazza, A.; Ragazzi, M. *Water Air Soil Pollut* **2017**, *228* (10), 388.

(12) Vandenbroucke, A. M.; Morent, R.; De Geyter, N.; Leys, C. *Journal of Hazardous Materials* **2011**, *195*, 30–54.

(13) Matsuda, S.; Kato, A. *Applied Catalysis* **1983**, *8* (2), 149–165.

(14) Iranpour, R.; Cox, H. H. J.; Deshusses, M. A.; Schroeder, E. D. *Environmental Progress* **2005**, *24* (3), 254–267.

(15) Bahri, M.; Haghighat, F.; Rohani, S.; Kazemian, H. *Chemical Engineering Journal* **2016**, *302*, 204–212.

(16) Preis, S.; Klauson, D.; Gregor, A. *Journal of Environmental Management* **2013**, *114*, 125–138.

(17) Faider, W.; Pasquiers, S.; Blin-Simiand, N.; Magne, L. *Plasma Sources Sci. Technol.* **2013**, *22* (6), 065010.

(18) Zheng, C.; Zhu, X.; Gao, X.; Liu, L.; Chang, Q.; Luo, Z.; Cen, K. *Journal of Industrial and Engineering Chemistry* **2014**, *20* (5), 2761–2768.

(19) Pasquiers, S. *The European Physical Journal - Applied Physics* **2004**, *28* (3), 319–324.

(20) Koeta, O.; Blin-Simiand, N.; Faider, W.; Pasquiers, S.; Bary, A.; Jorand, F. *Plasma Chem Plasma Process* **2012**, *32* (5), 991–1023.

(21) Pasquiers, S.; Blin-Simiand, N.; Magne, L. *Eur. Phys. J. Appl. Phys.* **2016**, *75* (2), 24703.

(22) Blin-Simiand, N.; Jorand, F.; Magne, L.; Pasquiers, S.; Postel, C.; Vacher, J.-R. *Plasma Chem Plasma Process* **2008**, *28* (4), 429–466.

(23) Munson, B. *Analytical Chemistry* **1971**, *43* (13), 28A – 43a.

(24) Blake, R. S.; Monks, P. S.; Ellis, A. M. *Chem. Rev.* **2009**, *109* (3), 861–896.

(25) Dehon, C.; Gaüzère, E.; Vaussier, J.; Heninger, M.; Tchapla, A.; Bleton, J.; Mestdagh, H. *Int. J. Mass Spec.* **2008**, *272* (1), 29–37.

(26) Dehon, C.; Lemaire, J.; Heninger, M.; Chaput, A.; Mestdagh, H. *Int. J. Mass Spec.* **2011**, *299* (2–3), 113–119.

(27) Spanel, P.; Smith, D. *Int. J. Mass Spec.* **1999**, *184* (2–3), 175–181.

(28) Spanel, P.; Smith, D. *Int. J. Mass Spec. Ion Proc.* **1997**, *167–168*, 375–388.

(29) Smith, D.; Chippendale, T. W. E.; Spanel, P. *Int. J. Mass Spec.* **2011**, *303* (2–3), 81–89.

(30) Sovová, K.; Dryahina, K.; Spanel, P. *IInt. J. Mass Spec.,* **2011**, *300 (1)*, 31-38

(31) Heninger, M.; Mestdagh, H.; Louarn, E.; Mauclaire, G.; Boissel, P.; Leprovost, J.; Bauchard, E.; Thomas, S.; Lemaire, J. *Anal. Chem.* **2018**, *90* (12), 7517–7525.

(32) Magne, L.; Pasquiers, S.; Blin-Simiand, N.; Postel, C. *J. Phys. D: Appl. Phys.* **2007**, *40* (10), 3112–3127.

(33) Mauclaire, G.; Lemaire, J.; Boissel, P.; Bellec, G.; Heninger, M. *Eur. J. Mass Spectrom.* **2004**, *10* (2), 155–162.

(34) Lemaire, J.; Thomas, S.; Lopes, A.; Louarn, E.; Mestdagh, H.; Latappy, H.; Leprovost, J.; Heninger, M. *Sensors* **2018**, *18* (5), 1415.

(35) Louarn, E.; Hamrouni, A.; Colbeau-Justin, C.; Bruschi, L.; Lemaire, J.; Heninger, M.; Mestdagh, H. *Int. J. Mass Spec.* **2013**, *353*, 26–35.

(36) Louarn, E.; Asri-Idlibi, A. M.; Leprovost, J.; Héninger, M.; Mestdagh, H. *Sensors* **2018**, *18* (12), 4252.

(37) Latappy, H.; Lemaire, J.; Heninger, M.; Louarn, E.; Bauchard, E.; Mestdagh, H. *Int. J. Mass Spec.* **2016**, *405*, 13–23.

(38) Le Vot, C.; Lemaire, J.; Pernot, P.; Heninger, M.; Mestdagh, H.; Louarn, E. *J Mass Spectrom* **2018**, *53* (4), 336–352.

(39) Moreau, N.; Pasquiers, S.; Blin-Simiand, N.; Magne, L.; Jorand, F.; Postel, C.; Vacher, J.-R. Propane. *J. Phys. D: Appl. Phys.* **2010**, *43* (28), 285201.

(40) Blin-Simiand, N.; Bournonville, B.; Coquery, P.; Jeanney, P.; Magne, L.; Pasquiers, S.; Tardiveau, P. Dissociation Kinetics of Acetone in a Sub-Atmospheric Pressure Nitrogen Plasma; Bratislava, Slovakia, 2016.

(41) St. John, W. M.; Estler, R. C.; Doering, J. P. *J. Chem. Phys.* **1974**, *61* (3), 763–767.

(42) Pastega, D. F.; Lange, E.; Ameixa, J.; Barbosa, A. S.; Blanco, F.; García, G.; Bettega, M. H. F.; Limão-Vieira, P.; Ferreira da Silva, F. *Phys. Rev. A* **2016**, *93* (3),

**SUPPLEMENTARY INFORMATIONS**

*Figure S1* Calibration signal of acetone for known introduced concentration of acetone (between 0 and 500 ppm). Introduced concentrations are calculated from partial pressure measurement using a MKS baratron® pressure gauge (MKS Instrument France, Le Bourget, France) calibrated to the atmosphere. Total pressure is 460mbar. The concentrations in BT4 apparatus are calculated according to Dehon et al[25].

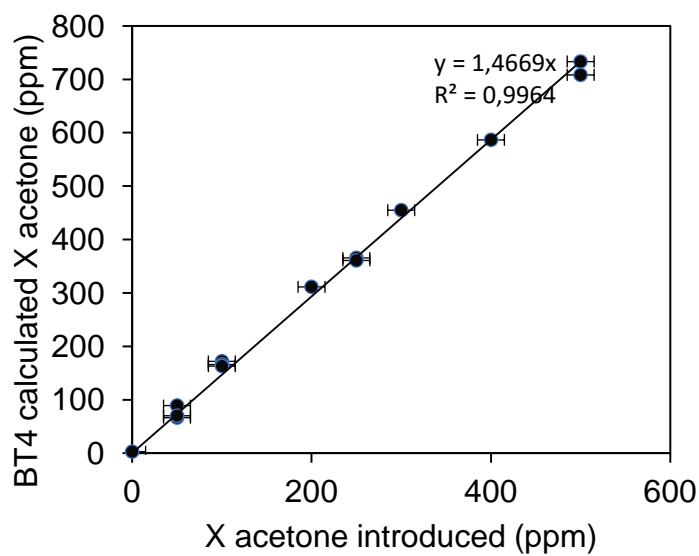

*Figure S2* Comparison of GC-FID (tr=1.56min) and BTrap/$O_2$ ($C_3H_8^+$) results for propane. Both curves present the same shape. The Btrap concentration is calculated for $k_M=1.36 \cdot 10^{-9}$ cm$^3$s$^{-1}$. The discrepancy at low $N_D$ is related to the FT treatment needed to deconvoluate the interferences at mass m/z 44 due to the first isotope of acetone fragment ($CH_3CO^+$).

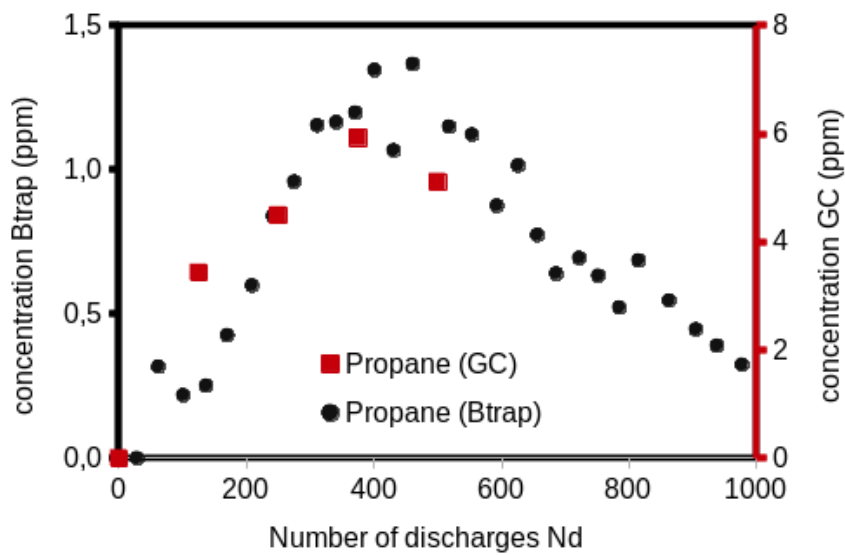

**Figure S3** a) Mass spectra at m/z 44 for $O_2^+$ precursor, 500ppm acetone, after circa 450 discharges. Five peaks are observed, three among them vary. Formulae are respectively: $CO_2^+$ (43.9898 u), $N_2O^+$ (44.0011 u), $^{13}C$ isotope of the $C_2H_3O^+$ fragment of acetone (44.0212 u), $C_2H_4O^+$ (44.0262 u) and $C_3H_8^+$ (44.0626 u). b) Signal evolution of the five masses identified. c) Mass value evolution during the experiment.

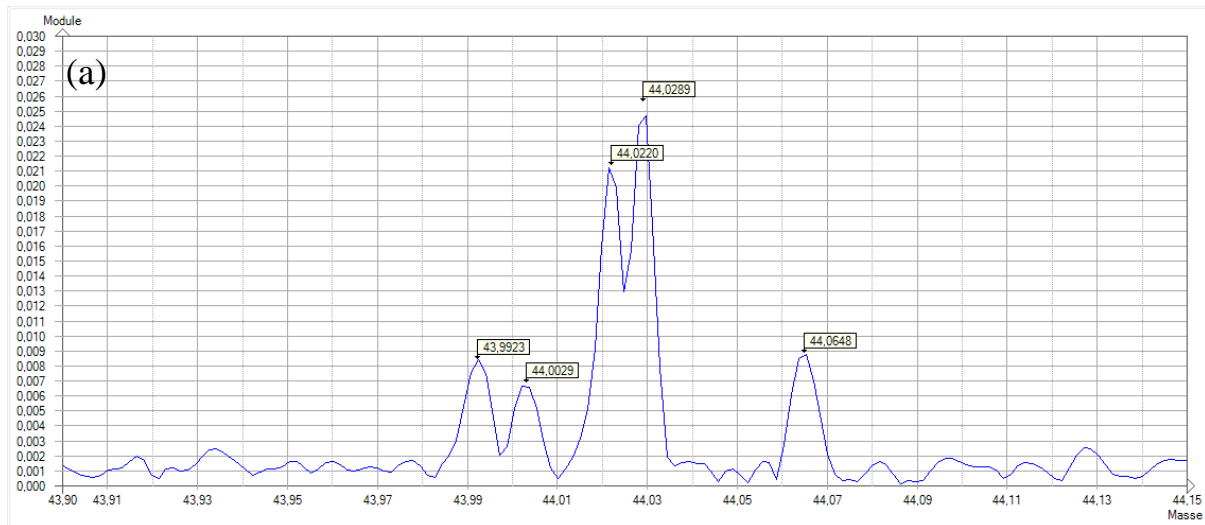
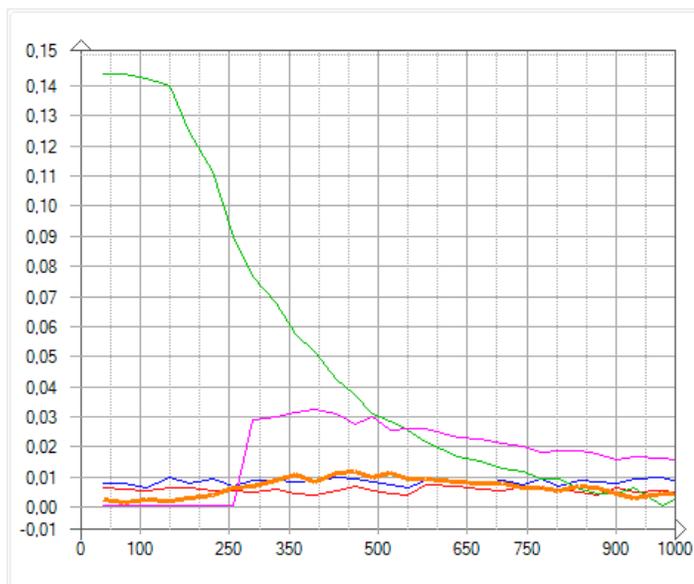
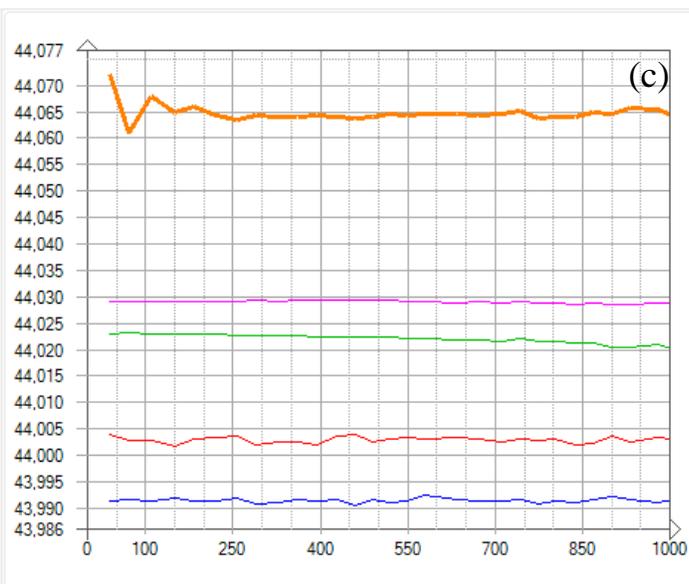